\documentstyle[aps,psfig,epsfig]{revtex}

\begin{document}

\title{Roughness Scaling in Cyclical Surface Growth}
\author{Subhadip Raychaudhuri$^{(1)}$, Yonathan Shapir$^{(1,2)}$, David
G. Foster$^{(2,3)}$, and Jacob Jorne$^{(2)}$}
\address{$^{(1)}$Department of Physics and Astronomy, University of
Rochester, Rochester, NY 14627 \\
$^{(2)}$Department of Chemical Engineering, University of
Rochester, Rochester, NY 14627 \\
$^{(3)}$Eastman Kodak Company, Rochester, NY14650}
\date{\today}
\maketitle

\begin{abstract}
    The scaling behavior of cyclical growth ({\it e.g.} cycles of alternating
deposition and desorption primary processes) is investigated
theoretically and probed experimentally. The scaling approach to
kinetic roughening is generalized to cyclical processes by
substituting the time by the number of cycles $n$. The roughness is
predicted to grow as $n^{\beta}$ where $\beta$ is the cyclical growth exponent.
The roughness saturates to a value which scales with the system size
$L$ as $L^{\alpha}$, where $\alpha$ is the cyclical roughness
exponent.
The relations between the cyclical exponents
and the corresponding exponents of the primary processes are
studied. Exact relations are found for cycles composed of primary
linear processes. An approximate renormalization group approach is
introduced to analyze non-linear effects in the primary processes.
The analytical results are backed by extensive numerical simulations
of different pairs of primary processes, both linear and non-linear.
Experimentally, silver surfaces are grown by a cyclical process
composed of electrodeposition followed by 50\% electrodissolution.
The roughness is found to increase as a power-law of $n$,
consistent with the scaling behavior anticipated theoretically.
Potential applications of cyclical scaling include accelerated testing of
rechargeable batteries, and improved chemotherapeutic treatment of
cancerous tumors.

\end{abstract}

\newpage

\section{INTRODUCTION}

From DLA to MBE, kinetic models of growth and aggregation have attracted
much attention \cite{family1} \cite{bara} \cite{villain} \cite{zhang}
\cite{meakin1} in the last
two decades due to diverse interests in physics, biology, chemistry and 
engineering. Kinetic roughening of non-equilibrium surface
growth were of particular interest.  Crystal growth, the growth of
bacterial colonies, and the formation of clouds in the upper
atmosphere \cite{pelletier} are
all examples of non-equilibrium phenomena which grow self-affine
rough surfaces.  On a fundamental level, surface growth problem is a
paradigm for a class of problems in non-equilibrium statistical mechanics.
One crucial aspect of
this class is the signature of scale invariance
and universality, very similar to those observed in the equilibrium critical
phenomena, or in non-linear dynamical systems \cite{burgers}.
Investigations focused primarily on the scaling behavior of the
surface roughness \cite{family1}, \cite{bara}, \cite{villain} \cite{zhang}
\cite{meakin1}.
More recently they have touched upon other aspects such as the distribution
of the surface width \cite{zia} \cite{plis}, distributions of the height and
the average height velocity \cite{zhang} \cite{bray} \cite{derrida}, 
density of extrema \cite{ziadasarma}, persistence \cite{PhysicsToday} 
and the maximal height \cite{subha}.

Processes which generates rough surfaces can be divided
into two classes: (i) growth processes ({\it e.g.} by deposition,
absorption) where the surface 
grows by adding material; and (ii) recession processes ({\it e.g.} by
erosion, dissolution, or desorption)
where material is being
taken off such that a rough surface is generated.
Examples of class (i)
include crystal growth \cite{family1} \cite{bara} \cite{villain}, 
electro-plating \cite{iwamoto1}, 
and biological growth 
\cite{family1} \cite{bara} \cite{meakin1} \cite{tumor}.
which have been studied widely.
Class (ii) includes chemical dissolution \cite{iwamoto2}, and
received much less attention than the first class.

Self-affine surfaces generated by growth (or recession)
can be described using scaling analysis
of the surface roughness.
In general, the surface is
characterized by the height $h(\vec{r},t)$ appropriate to the
($\tilde d = d-1$) dimensional substrate
of linear size L.  The width of the surface $W(L,t)$ at a time
$t$ is given by
\begin{equation} \label{eq:}
 W(L,t) = \Bigl \langle \, \Bigl( h( \vec{r},t) - \langle h(\vec{r},t) 
   \rangle \Bigr)^{2} \,  \Bigr \rangle   ^{\frac{1}{2}},
\end{equation}
where $ \langle h(\vec{r},t) \rangle$ is the average height which is linear in time
$\langle h(\vec{r},t) \rangle = vt$, 
$v$ being the average growth (or recession)
velocity. The angular bra-ket $\langle \rangle$ denotes average 
over both lateral
sites and the ensemble of surface configurations. 
 
$W(L,t)$ scales as \cite{family2}
\begin{equation} \label{eq:}
 W(L,t) \sim  L^{\alpha} \,  f(L/ \xi(t))
        \sim  L^{\alpha} \,  f(L/ t^{1/z}),
\end{equation}
where $\xi(t) \sim t^{1/z}$ is the lateral correlation length
and $f$ is the scaling function.
\newline For large time \hspace{0.1in}
$t \gg L^{z}$: \hspace{0.4in} $W \sim L^{\alpha}$,
\newline while for \hspace{0.5in}
$t \ll L^{z}$:  \hspace{0.4in} $W \sim t^{\beta}$,
\newline where $\beta = \alpha /z$ is the growth exponent.
The height difference
correlation function
\begin{equation} \label{eq:}
\Delta (r,t)= \Big \langle \Big( h(\vec{r}_{o}+\vec{r},t)
              -h(\vec{r}_{o},t) \Big)^{2} \Big \rangle,
\end{equation}
obeys scaling similar to that of the roughness

\begin{equation} \label{eq:}
 \Delta (r,t) \sim  r^{2 \alpha} f(r/ \xi(t))
              \sim  r^{2 \alpha} f(r/ t^{1/z}).
\end{equation}
For \hspace{0.35in}
$r \ll t^{1/z}$: \hspace{0.4in} $\Delta \sim r^{2 \alpha}$,
\newline while for \hspace{0.075in}
$r \gg t^{1/z}$:  \hspace{0.35in} $\Delta \sim t^{2 \beta}$.
\newline The relation between the height-difference
correlation function $\Delta(r,t)$ defined above
and the equal-time height-height correlation function
$ C(r,t) = \bigl \langle h(\vec{r}_{o}+ \vec{r},t) \, 
h(\vec{r}_{o},t) \bigr \rangle$
is simple
\begin{equation} \label{eq:}
      \Delta (r,t) =  2 \, [ C(0,t) - C(r,t)].
\end{equation}
Dynamic Scaling in Fourier space is easily studied in terms
of the structure factor $S(\vec{q},t) = \bigl \langle h(\vec{q},t)
\, h(\vec{-q},t) \bigr \rangle$,
where $h(\vec{q},t)$ is the Fourier transform of the height
$h(\vec{r},t)$. The scaling hypothesis Eq. (2) can be translated
to the Fourier space such that
\begin{equation} \label{eq:}
   S(\vec{q},t) = q^{-d -2\alpha} \, g(q/t^{-1/z}).
\end{equation}
Surface width $W$ can be readily calculated from $S(\vec{q},t)$
using the relation $W^{2}(L,t) = (1/L^{\tilde d}) \sum_{\vec{q}} 
S(\vec{q},t)$.
In the theoretical analysis of surface growth it is convenient to
work in Fourier space and compute first the structure
factor rather than the roughness $W$ itself. Also, in some experiments
the surface is probed by scattering processes which provide its
structure factor.

All processes within
the same
universality class share the same critical exponents.
Their continuum growth equations
differ at most by terms which are rendered irrelevant by the
renormalization group flow. The asymptotic continuum
stochastic equations corresponding to different
universality classes (indexed by $i = 1,2...$)
are of the form of a Langevin equation
\begin{equation} \label{eq:}
  \frac{\partial h(\vec{r},t)}{\partial t} = A_{i} \{h\} +
\eta_{i}(\vec{r},t) + v_{i},
\end{equation}
where $A_{i}$\{$h\}$ is a local functional depending on the spatial
derivatives of $h(\vec{r},t)$ and the noise $\eta_{i}(\vec{r},t)$
reflects the the random fluctuations in the deposition process 
and satisfies $\langle \eta_{i}(\vec{r},t) \rangle = 0$ and
$ \langle \eta_{i}(\vec{r}_{1},t_{1}) \, \eta_{i}(\vec{r}_{2},t_{2}) 
\rangle = 2D_{i}
\delta ^{\tilde d} (\vec{r}_{1}-\vec{r}_{2}) \, \delta(t_{1}-t_{2})$. Specific
generic processes and their
universality classes are reviewed later (Sec. III).

In the present paper we focus on cyclical processes in which deposition
and desorption
are occurring alternatingly. Examples of cyclical
processes are abundant in nature and technological applications.
A technologically important
example are rechargeable batteries, where metal is electrodeposited on an
electrode during the discharge, followed by partial electrochemical
dissolution of this metal during recharging. In chemotherapeutic treatment
of cancer the malignant cells are subjected to a recessive cyclical
process. 

The basic premise of our scaling approach to cyclical processes,
is that the number of cycles $n$ should substitute for the time variable.
So we propose for cyclical processes the
scaling relation of Eq. (1) be replaced with
\begin{equation} \label{eq:}
W_{c}(L,n) \sim L^{\alpha} \, f_{c}(L/ \xi_{c}(t))
           \sim L^{\alpha} \, f_{c}(L/n^{1/z}),
\end{equation}
where $\xi_{c} \sim n^{1/z}$ is the correlation length and $f_{c}$ is
the cyclical scaling function.
\newline For large n \hspace{0.1in}
$n \gg L^{z}$: \hspace{0.4in} $W \sim L^{\alpha}$,
\newline while for \hspace{0.25in}
$n \ll L^{z}$:  \hspace{0.4in} $W \sim n^{\beta}$,
\newline where $\beta = \alpha /z$ is the growth exponent.

The rest of the paper is organized as follows:
The next section II is devoted to our analytical analysis.
We will prove that the equation (8) for
cyclical scaling holds asymptotically for linear processes.
In addition we will show how to obtain
the scaling exponents of the cyclic process,
given those of the two primary processes.
An approximate RG approach is introduced to study non-linear effects.
Sec. III contains a brief review of the generic universality
classes and their lattice algorithms. Then we introduce new algorithms
to implement numerically different recession processes.
Actual results of our simulation of cyclical
processes are described in Sec. IV. In Sec. V the experimental
findings from cyclical electrodeposition/dissolution of silver are discussed.
Sec. VI contains the conclusions with a discussion of potential
practical applications of the cyclical scaling approach. A short summary 
of some of the results was published in Ref. \cite{shapir}.

\section{ANALYTICAL RESULTS}

The analytical approaches are all based on the stochastic equation, like
Eq. (7),
which describes the growth process.
Thus, we begin by obtaining the stochastic equation of the cyclical process.
The first primary process is denoted by $i=1$ and the second
by $i=2$. The durations of the first and the second processes are
$T_{1}=pT$ and $T_{2}=(1-p)T$, respectively (p and 1-p are the fractional
parts of both processes). The total time period for
one cycle is $T = T_{1} + T_{2}$. The cyclic growth equation
in terms of the basic two processes can be expressed as
\begin{equation}\label{eq:}
\frac{\partial h}{\partial t}=[a_{1}h + \eta _{1}
 + v_{1}]\Theta(p-f(t)) +
[a_{2}h + \eta_{2} + v_{2}]\Theta(f(t)-p),
\end{equation}
where $f(t)$ is defined as the fractional part of $t/T$
and $\Theta(x)$ is the unit step function.

\subsection{Linear Primary Processes - Exact Calculation of 
$S_{c}(\vec{q},t)$}

First we will
consider cyclical processes for which both the primary processes are
linear. For linear processes $A_{i}\{h\} =
a_{i}(\vec{\nabla})h(\vec{r},t)$, where $a_{i} (\vec{\nabla})$
is a linear differential operator. The time reversal symmetry is
obeyed in this case if the height is measured relative to the
average height. The average height indeed depends on the
growth ($v_{i} > 0$) and the recession ($v_{i} < 0$) nature of the
primary processes. In terms of the average
velocity $v_{c} = pv_{1}+(1-p)v_{2}$, it is given by

\begin{equation} \label{eq:}
\langle h(\vec{r},t) \rangle /T= nv_{c} +v_{1} \, f(t) \, \Theta 
\big( p-f(t) \big) +
[(v_{1}-v_{2})p+v_{2} \big( f(t) \big)] \, \Theta \big( f(t)-p \big).
\end{equation}

The roughness is insensitive to the sign of $v_{i}$ and
hence will not distinguish
between a growth/growth and a growth/recession cyclical processes
as long as the basic processes remain linear.

For linear processes Langevin equations of the form
of Eq. (7) are easily solved in Fourier space to yield  
(assuming spatial isotropy in the basal plane) the structure factor

\begin{equation} \label{eq:}
S(q,t) = \exp$\{$-2a(q)t$\}$ S(q,0) +
         \frac{D}{a(q)}[1-\exp(-2a(q)t)],
\end{equation}

where $S(q,0)$ is the structure factor at $t=0$ which
contains the information of the initial roughness. 
During the $nth$ cycle of the cyclical growth,
the structure factor $S_{c}(q,\big((n-1)+p)T \big)$ generated
by the first primary process (of duration$T_{1}= pT$)
is assigned as the initial condition for the second primary process.
The second process lasts for $T_{2}=(1-p)T$ to yield the
structure factor $S_{c}(q,nT)$ of the cyclical process after $n$
cycles. This is again used as the initial structure factor
for the first process in the $(n+1)th$ cycle.

 During the first cycle,
the structure factor after the completion of the
first primary process becomes 
\begin{equation} \label{eq:}
S(q,T_{1}) = \exp$\{$-2\bar a_{1}$\}$ S(q,0) +
         \frac{D_{1}}{a_{1}}[1-\exp(-2\bar a_{1})],
\end{equation}
where $\bar{a_{i}} = a_{i}(q)T_{i}$ was defined.
This is the initial structure factor for the second primary process
and after the first complete cycle we obtain

\begin{eqnarray}
S(q,T_{2}) & = & \exp \{ -2(\bar a_{1}+\bar a_{2} \} \,
S(q,0)     \nonumber \\
& & + \exp \{ -2\bar a_{2} \}  \Bigl \{ \frac{D_{1}}{a_{1}}
[1-\exp(-2\bar a_{1})] \Bigr \}
       +  \frac{D_{2}}{a_{2}}[1-\exp(-2\bar a_{2})],
\end{eqnarray}

where $S_{c}(q,T) \equiv S(q,T_{2})$.
Proceeding in this manner, we finally arrive at the structure factor
$S_{c}(q,n) \equiv S(q,nT)$ of the cyclic growth after
$n$ cycles as a geometric series
which can be readily summed to yield

\begin{eqnarray}
S_{c}(q,n) & = & \exp \{ -2\bar{a}_{c}n \}  S(q,0) \nonumber \\ 
& &  +
\left[\frac{D_{1}}{a_{1}}\exp(-2\bar a_{2})
\{ 1-\exp(-2\bar a_{1}) \}
+\frac{D_{2}}{a_{2}} \{1-\exp(-2\bar{a}_{2})\}\right]
\left[{1-\exp(-2\bar{a}_{c}n)\over1-\exp(-2 \bar{a}_{c})}\right],
\end{eqnarray}

where $\bar a_{c} = a_{c}T$ with

\begin{equation} \label{eq:}
         a_{c} = [a_{1}p + a_{2}(1-p)].
\end{equation}

    In the scaling limit of small $q$, Eq. 14 for $S_{c}(q,n)$ reduces to
\begin{equation} \label{eq:}
S_{c}(q,n) \sim
\frac{D_{c}}{a_{c}(q)}
\left[1-\exp \big(-2a_{c}(q) \, T \, n \big) \right],
\end{equation}
with the effective noise strength for the cyclic process is 
defined as
 
\begin{equation} \label{eq:}
      D_{c}=pD_{1}+(1-p)D_{2}.
\end{equation}
 
In terms of the effective parameters
$a_{c}$ and $D_{c}$, the above structure factor for the cyclical process
resembles that of a generic linear growth process (see Eq. (11)
with the number of cycles $n$ as the new time variable.
Effectively the time is being coarse-grained over a period
of $T$ by eliminating the high frequency $( > 2\pi/T)$ modes.
Hence we can use the standard scaling analysis (Eq. (6))
of the structure factor to determine the scaling exponents in the case
of a cyclical growth.

     After a large number of cycles $n$
the interface width becomes saturated. In that limit, Eq. (15)
becomes  $S_{c}(q,n) \sim \frac{D_{c}}{a_{c}(q)}$. The roughness
exponent of the cyclic process is determined by the $q \rightarrow 0$
divergence of $\frac{1}{a_{c}(q)}$. Since
$a_{i}(q) \sim q^{z_{i}}$, it is the
process with smaller $z_{i}$ which dominates the asymptotic
cyclical roughness, and the roughness exponent is given by
$\alpha_{c}
= min(\alpha_{1}, \alpha_{2}) = \frac{1}{2} 
\{ min(z_{1}, z_{2}) - (d-1) \}$. The primary
process with the {\it smaller}
roughness imposes its roughness exponent on the combined
cyclical  process. The larger $\alpha_{i}$ appears as a correction to the
scaling exponent. Whether or not it is the leading one depends on how its
contribution compares with that of the subleading term in $a_{i}(q)$ of
the dominating primary process.
Note that a subleading term in $a_{c}(q)$ might affect the behavior
on a smaller scale, if its amplitude is large. In that case, the leading
behavior takes over only beyond a crossover length (at which both
contributions are comparable). Since the amplitudes of $a_{1}$ and 
$a_{2}$ in $S_{c}(q,n)$ are proportional
to $p$ and $(1-p)$, respectively,
the longer the non-dominant process lasts, the larger is the
crossover scale, as could be expected.
However, although in $a_{c}(q)$ only the dominant $a_{i}(q)$ is
important beyond this crossover scale, this is not the case for
the effective noise-correlator $D_{c}$,
which is a scale independent constant in Eq. (16). Thus, the {\it amplitude} of
the leading power-law roughness is determined by both the
primary processes.

     In the growing phase of the interface roughness, the dynamic exponent
$z$ will dictate the cyclical power law behavior. Since $n$ is
multiplied by $a_{c} = [a_{1}p + a_{2}(1-p)]T$ in Eq. 16 and
$a_{i}(q) \sim q^{z_{i}}$, the process with smaller $z_{i}$ will
again dominate the cyclical dynamics in the $q \rightarrow 0$ asymptotic
limit to yield $z_{c}=min(z_{1},z_{2})$.
Then for the initial cycles ($nT \ll L^{z}$),
the surface width will scale as $n^{\beta}$, with $\beta_{c}= \alpha_{c}/
z_{c}$. In essence, the scaling exponents of the cyclic growth will
be identical to those of the primary process with the smaller $z_{i}$.

\subsection{Coarse-Graining Approach: The Cyclic Propagator 
$G_{c}(\vec{r},t)$}

In the previous section the structure factor $S_{c}(\vec{q},t)$ 
(and hence its Fourier Transform (FT) $S_{c}(\vec{r},t)$) was obtained
exactly. The same method of successive integration of the cyclical 
equation of motion may be applied to derive the cyclic propagator
$G_{c}(\vec{r},t)$ (or its FT $G_{c}(\vec{q},t)$). However, the
expression is cumbersome and not very useful. Since we only need 
the long-time behavior, we will take a different route of 
coarse-graining the equation of motion such that the equation for
time scales larger than $T$ will be derived and $G_{c}(\vec{r},t)$ 
could be read 
from it. This will also provide a direct connection to the RG approach
introduced in the next section to study the behavior of non-linear 
systems. 

Our starting point is Eq. (9), which we choose to integrate over one 
cycle, such that the remaining equation will be a difference equation
between the average heights of the consecutive cycles. So assume we average
Eq. (9) on time $t \in [nT, (n+1)T]$. The integral will be divided 
into integration over the two intervals: interval $(1)$ 
$[nT, nT+T_{1}]$ and interval $(2)$ $[nT+T_{1},(n+1)T]$. We 
define the average height in the $n^{th}$ cycle as 

\begin{equation} \label{eq:}
 H(\vec{r},n) = \frac{1}{T} \int_{n}^{n+1} dt \, h(\vec{r},t).
\end{equation}

Our goal is to obtain the equation of motion for $H(\vec{r},n)$
on time scales $t > T$. The FT of Eq. (9) is

\begin{eqnarray}
\frac{\partial h(\vec{q},t)}{\partial t} & = & 
[a_{1}(\vec{q}) \, h(\vec{q},t)+\eta_{1}+v_{1}]\Theta(p-f(t))
+ [a_{2}(\vec{q}) \, h(\vec{q},t)+\eta_{2}+v_{2}]\Theta(f(t)-p) \nonumber \\
 & = & a_{c}(\vec{q}) h(\vec{q},t)  
+ [\Delta a_{1}(\vec{q}) \, \Theta(p-f(t))
+ \Delta a_{2}(\vec{q}) \, \Theta(f(t)-p)] \, h(\vec{q},t) \nonumber \\
& & + (\eta_{1}+v_{1}) \, \Theta(p-f(t)) +(\eta_{2}+v_{2}) \, \Theta(f(t-p)), 
\end{eqnarray}

where $\Delta a_{1}(\vec{q}) = \Big( a_{1}(\vec{q})-a_{c}(\vec{q})
\Big)$ and  $\Delta a_{2}(\vec{q}) = \Big( a_{2}(\vec{q})-a_{c}
(\vec{q}) \Big)$.
Let us now integrate over one cycle each of the terms, beginning 
with the {\it l.h.s.}

\begin{eqnarray}
\frac{1}{T} \int_{n}^{n+1} dt \, \frac{\partial h (\vec{q},t)}{\partial t}
& = & \frac{1}{T} \, [h(\vec{q},(n+1)T) - h(\vec{q},nT)] \nonumber \\
& = & \frac{1}{T} \, [H(\vec{q},n+1) - H(\vec{q},n)] \nonumber \\
& & \quad + \frac{1}{T}  \, 
\Bigl \{ \Bigl( h(\vec{q},(n+1)T) - H(\vec{q},n+1) \Bigr)
               - \Big( h(\vec{q},nT) - H(\vec{q},n) \Bigr) \Bigr \}.
\end{eqnarray}

The first term can be expressed as  
$ \frac{\Delta H(\vec{q},n)}{T \Delta n }$. The
second term contains the differences between the height at the 
beginning of the cycle and its average over a cycle. This term thus
reflects the behavior within each cycle and is thus irrelevant to the
behavior on the coarse grained scale and will be dropped. (Note that
it will vanish asymptotically and it is irrelevant since in the 
continuum limit it features a second derivative with respect to $t$).

The integration of the first term on the right-hand side yields

\begin{equation} \label{eq:}
\frac{1}{T} \int_{n}^{n+1} a_{c}(\vec{q}) \, h(\vec{q},t) \, dt
 = a_{c}(\vec{q}) H(\vec{q},n).
\end{equation}

Integrating the second term

\begin{eqnarray}
& & \frac{1}{T} \int_{n}^{n+1}  \,[\Delta a_{1}(\vec{q}) \, \Theta(p-f(t))
+ \Delta a_{2}(\vec{q}) \, \Theta(f(t)-p)] \, h(\vec{q},t) 
\, dt \nonumber \\
 & = & \frac{\Delta a_{1}(\vec{q})}{T} \int_{n}^{n+p} h(\vec{q},t) \, dt 
+ \frac{\Delta a_{2}(\vec{q})}{T} \int_{n+p}^{n+1} h(\vec{q},t) \, dt 
\nonumber \\ 
 & = &  \Delta a_{1}(\vec{q}) \, p \,  
\Bigl  \{ \frac{1}{T_{1}} \int_{n}^{n+p} h(\vec{q},t) dt \Bigr \}
+ \Delta a_{2}(\vec{q}) (1-p) 
\Bigl  \{ \frac{1}{T_{2}} \int_{n+p}^{n+1} h(\vec{q},t) dt \Bigr \}
\nonumber \\
 & = & \Delta a_{1}(\vec{q}) \, p \, H_{1}(\vec{q},n) 
 + \Delta a_{2}(\vec{q}) \, (1-p) \, H_{2}(\vec{q},n),  
\nonumber \\
& = & [\Delta a_{1}(\vec{q})p  + \Delta a_{2}(\vec{q})(1-p)] 
 H(\vec{q},n)  \nonumber \\
& & \qquad + \{ \Delta a_{1}(\vec{q})p [H_{1}(\vec{q},n) - H(\vec{q},n)]
 +   \Delta a_{2}(\vec{q})(1-p) [H_{2}(\vec{q},n) - H(\vec{q},n)] \},
\end{eqnarray}

where $H_{1}(\vec{q},n)$ and $H_{2}(\vec{q},n)$ are the average
heights during the two primary processes respectively 
(on the $n^{th}$ cycle). Upon coarse-graining

\begin{eqnarray}
& & \frac{1}{T} \int_{n}^{n+1} [\Delta a_{1}(\vec{q}) \, \Theta(p-f(t))
+ \Delta a_{2}(\vec{q}) \, \Theta(f(t)-p)] \, h(\vec{q},t) \, 
dt  \nonumber \\
& & \qquad = [\Delta a_{1}(\vec{q})p  + \Delta a_{2}(\vec{q})(1-p)] 
\, H(\vec{q},n)
\end{eqnarray}

This term contains information on the structure factor within the cycle
and must vanish in the coarse-grained equation, since $a_{c}$ was chosen 
(Eq. (15)) such that

\begin{equation} \label{eq:}
  (a_{1} - a_{c})p + (a_{2} - a_{c})(1-p) = 0.
\end{equation}

The velocity term yields the average velocity

\begin{equation} \label{eq:}
 \frac{1}{T} \int_{n}^{n+1} dt (v_{1} \, \Theta(p-f(t)) + 
  v_{2} \, \Theta(f(t)-p))  = v_{c}.
\end{equation}

The noise $\eta$'s are random variables and we define its 
coarse-grained term by

\begin{equation} \label{eq:}
 \eta_{c}(n) = \frac{1}{\sqrt{T}} [\int_{n}^{n+p} dt \, \eta_{1}(t)
                           + \int_{n+p}^{n+1} dt \, \eta_{2}(t)].   
\end{equation}

It obeys
$ \langle \eta_{c}(n) \rangle = 0$ and $ \langle \eta_{c}(n)
\, \eta_{c}(m) \rangle = 2D_{c} \delta_{n,m}$, and $D_{c}$ 
satisfies Eq. (17).

Collecting all the terms, the coarse-grained difference equation
for $H(\vec{q},n)$ is

\begin{equation} \label{eq:}
 \frac{\Delta H(\vec{q},n)}{T \Delta n } = a_{c}(\vec{q})H(\vec{q},n)
               + \eta_{c}(\vec{q},n) + v_{c}.
\end{equation}

For $n \gg 1$, the difference equation is equivalent to the corresponding
differential equation $ \frac{\Delta H}{\Delta n} \simeq 
\frac{\partial H}{\partial n} $. 

The cyclical propagator may be read from this linear equation

\begin{equation} \label{eq:}
 G_{c}(\vec{q},n-m) = \exp \{- a_{c}(\vec{q})(n-m)\} \Theta(n-m).
\end{equation}

This propagator obeys the usual relation to the structure factor if 
$n$ is treated as a continuous variable replacing the time $t$ 

\begin{equation} \label{eq:}
  S_{c}(\vec{q},n) = \frac{D_{c}}{a_{c}(\vec{q})}[1- |G_{c}(\vec{q},n)|^{2}].
\end{equation}

\subsection{Non-Linear Primary Processes}

Stochastic non-linear equations of that type can be analyzed using
the perturbative dynamic RG approach \cite{bara} \cite{kardar}. We
develop an approximate RG technique for cyclical processes with one
or both the primary processes being non-linear.
The first
step is to set aside (for the initial RG iterations) all the
non-linearities from the primary processes
and take only the linear Langevin equations. It was shown above how
to solve for the structure factor of such cyclical processes
and get an effective $S_{c}(q,n)$ (see Eq. (16)) of a non-cyclic process.
Similarly we have shown in the previous subsection how to obtain the
propagator $G_{c}(q,n)$ of the same effective linear, non-cyclical process.
The second step is to take the effective ``free'' propagator $G_{c}(q,n)$
and then add back all the non-linear terms of the primary processes as
perturbations. The bare couplings of the non-linear terms are multiplied
by $p$ and $(1-p)$ to take into account the relative durations
of the two primary processes. The third and the final step is to study
the RG flow of the parameters and determine the fixed points
of the transformations and hence the scaling exponents. This dynamic
RG procedure is approximate in the sense that the initial flow of the
couplings is shifted, but as long as the starting point is
not close to a separatrix in the parameter space ({\it i.e.} a
border line between basins of
attraction of two different fixed points)
this will not alter the ultimate fixed-point
of the RG flow.

The cyclical process might have only one non-linear primary process
with the non-linear perturbation being relevant under RG
transformations. Then the cyclical scaling exponents are given by
the exponents of the non-linear primary process.
Otherwise, the non-linear term turns out to be irrelevant with respect to
the dominant linear term (present in the coarse grained
cyclic free propagator) of the other (linear) primary process, which force
its own scaling exponents for the cyclic growth.

Both the primary processes might contain non-linear terms which are
treated as perturbations to the free propagator as described above.
In most cases, one of the non-linear term will make the other one
irrelevant under the RG flow and the primary process containing the
relevant non-linearity will carry through its scaling exponents
for the cyclical process. If both the non-linear terms are relevant,
then the possibility of new cyclical growth exponents (different than
those of both primary processes) can not be ruled out. A prime candidate
for such a behavior will be a cyclical process in $d = 2+1$ in which
one process has a relevant non-linearity while the second process
has a relevant anisotropy in the basal plane \cite{dkim}

\section{GROWTH MODELS AND UNIVERSALITY CLASSES}

In this section We present a brief review of the basic models of 
kinetic growth.
The discrete lattice models for 
simulations corresponding to each universality 
class are presented. We  were
able to generalize most of the lattice models for  desorption processes
(reverse of growth) to  be used in absorption/desorption cycles. These
new desorption algorithms are described in detail.

$1)$ ${\bf Random \; Deposition \; (RD)}$ - This is simplest of all 
possible growth processes. From a 
randomly chosen site from the surface, a particle falls vertically until
it
reaches the top of the column under it, whereupon it is deposited. In this
case
$$                     
                   A_{RD}=0.
$$
Scaling exponents are: $\beta=0.5$ in all dimensions but $\alpha$ is
not defined for this model because the interface is never saturated.

RD Algorithm - A column i is chosen (d=1+1) randomly and its height
h(i,t) is increased one. For the reverse process we just decrease the
height h(i,t) of an arbitrarily chosen site $i$ by one. 

$2)$ ${\bf Edwards-Wilkinson \; (EW) \; Universality}$ -  
In this model we have surface relaxation (in addition
to the random deposition) which is introduced by the term 
$$
                  A_{EW}= \nu \nabla^{2} h.
$$
We can exactly calculate the scaling exponents: $\alpha=\frac{3-d}{2}$, 
$\beta=\frac{3-d}{4}$, and $z=2$.
In EW growth \cite{ew} a randomly deposited particle 
can diffuse along the surface up to a 
a finite distance and sticks to a local height minimum. Due to this 
relaxation, surface becomes smooth compared to the random growth model
and finally  the interface roughness is saturated because of the
correlations among the neighboring heights, to a value 
$ \sim L^{\alpha}$, where $L$ is the lateral size.  

Family Model - This model was introduced by Family \cite{family3} 
to simulate EW growth. A particle is dropped on a randomly chosen column i 
(in $d = 1+1$) and sticks to the top of the column i, i+1 or i-1, 
depending on which of the three columns has the smallest height.
To simulate a desorption
process  (with EW exponents), the height $h(i,t)$ of the site
$i$ is compared to the
heights $h(i-1,t)$ and $h(i+1,t)$ of the neighboring sites. Then we simply
take a particle off from
the column i, i+1 or i-1, depending on which of the three columns has the
largest height. If case of a tie involving
the site i we take out a particle from that site, otherwise the tie is
broken
randomly with equal probability. The whole process can be thought as 
desorption with surface relaxation. The relaxation length is restricted to
the nearest neighbors because the scaling exponents are independent of
the relaxation length. In Fig. 1 the roughness $W$ is plotted against
time on a log scale for increasing system sizes. The slope of
the straight line fitted to the early time roughness yields the growth
exponent $\beta = 0.23 \pm 0.02$. The roughness exponent
$\alpha = 0.48 \pm 0.03$ is extracted from the dependence of the
saturation roughness on the linear size of the system 
(Inset of Fig. 1).  
These values of the scaling exponents are in agreement with the
corresponding EW values. Indeed,
the surface roughness for the desorption
process shows (Fig. 1) very similar behavior to that obtained 
for the growth process \cite{family3}.   
So the discrete rules described above can be treated as a valid
algorithm for EW desorption. 

$3)$ ${\bf Kardar-Parisi-Zhang \; (KPZ) \; Universality}$ - 
This describes growth in a direction locally normal 
to the interface \cite{kardar}. 
Its leading effect is to add a non-linear term to the 
EW surface relaxation term
$$
       A_{KPZ}= \nu \nabla^{2} h + \frac{\lambda}{2} (\nabla h)^{2}.
$$
Scaling exponents for KPZ equation are exactly known in $d=1+1$:
$\alpha=1/2$, $\beta=1/3$, and $z=3/2$.  In $d = 2+1$ approximate (from
numerical simulations) values for the exponents are: $\alpha \simeq 0.39$,
$\beta \simeq 0.24$ and $z \simeq 1.61$. KPZ can be simulated using
different atomistic growth algorithms of which two we will describe
because we generalized those to the case of desorption and will be used
in our simulations of cyclical growth.

Ballistic Deposition (BD) Model - A particle is released from 
a randomly chosen position above the
surface, located at a distance larger than the maximum height of the
interface. The particle follows a straight vertical path until it reaches
the surface, whereupon it sticks \cite{vold} \cite{meakin2}. If h(i,t)
is the height of the column i (chosen randomly in $d = 1=1$) at time
$t$ then the BD growth rule is:
h(i,t+1)=max[h(i-1,t), h(i,t)+1, h(i+1,t)]. For the reverse process the
algorithm will be changed to 
h(i,t+1)=min[h(i-1,t), h(i,t)-1, h(i+1,t)].
Although physically unrealistic (contrary to its growth counterpart) this
desorption model is formally the "anti-BD" process.
  
Restricted-Solid-On-Solid (RSOS) Model - 
This algorithm (also known as KK model) introduced by Kim and 
Kosterlitz \cite{kost} 
gives KPZ exponents and known to yield reliable results even for
small system sizes. The growth rule is to randomly select a site on a cubic
(d-1) dimensional lattice and to permit growth by letting the height
of the interface $h_{i} \rightarrow h_{i}+1$ provided the restricted
solid-on-solid restriction on neighboring heights $|\Delta h|$ =
$0,1,....,N$ ($N \ge 1$) is obeyed at all stages. In a similar way we 
can simulate an erosion process where we decrease the height 
($h_{i} \rightarrow h_{i} - 1$) of the site $i$ provided the RSOS
condition $|\Delta h|$ = $0,1,....,N$ is satisfied. In Fig. 2 we
show the results of simulations using this desorption rule. The values
of the scaling exponents ($\beta = 0.33 \pm 0.02$ and $\alpha =
0.52 \pm 0.02$) obtained are consistent with the corresponding 
KPZ values.

The RSOS deposition or desorption rule described above 
is defined starting from a flat interface at $t=0$. In a cyclical
surface growth model with a primary process not obeying the RSOS
condition can destroy the height difference restriction of the RSOS model.
We can still use a growth rule similar to the RSOS model described above, 
which seems to be behaving like a KPZ growth (as far as the scaling
exponents are concerned). In this extended model of RSOS, we choose a
site $i$ randomly and add a particle on it only if the height of
that site is less than or equal to the heights of the neighboring
sites (which would give us the normal RSOS with $|\Delta h| \, = 0,1$
starting from a flat interface). For the reverse process a 
particle is taken off from a site only if the height of that site is
greater than or equal to the heights of the neighboring sites.

$4)$ ${\bf Mullins-Herring \; (MH) \; Universality}$ - 
In conserved growth situations where "surface diffusion"
is dynamically significant in the absence of any 
EW relaxation process, the
growth process may belong to the MH universality class (also known as
Das Sarma-Tamborenea (DT) or Wolf-Villain (WV) class)
\cite{MH},\cite{dasarma1},\cite{wolf}.
The linear surface diffusion equation for MH has a term 
$$               
                        A_{MH}= -K \nabla^{4} h.
$$
The critical exponents for the MH growth universality are exactly known 
theoretically: $\alpha=(5-d)/2$, $\beta=(5-d)/8$ and $z=4$. In one
dimension 
(d=1+1), the roughness exponent, $\alpha=1.5$, exceeds unity, implying
that
the large scale steady state  morphology of the growing interface is not
self-affine in d=1+1. The issue of whether the MH universality is only a 
crossover phenomenon or a true universality class is still 
debated \cite{dasarma2}. There
are two lattice models to simulate MH growth.
 
Das Sarma-Tamborenea (DT) Model - In this model \cite{dasarma1} a particle, 
after being deposited on a randomly chosen site,
relaxes only to a nearest kink site i.e. it seeks only to increase the 
number of neighbors.

Kim-Das Sarma (KD) Model - Choose a site i randomly and add a 
particle on i-1, i or i+1 (in
d=1+1), depending on which site has a larger curvature {$\nabla^{2} h =
[h(x+1) +h(x-1) -2h(x)$] than that at the site i \cite{kim}. 
If there is more than one 
satisfying the larger curvature condition, we choose one among them
randomly.
If the curvature of the site i is one of the larger curvatures, we add the 
particle at site i.

$5)$ ${\bf Molecular \, Beam \, Epitaxy \; (MBE) \;  Universality}$ - 
The most relevant universality class in the
context of
conserved epitaxial growth is the Molecular Beam Epitaxy (MBE)
universality
(also goes by the name Lai-Das Sarma-Villain (LDV) universality) 
\cite{laid} \cite{vill}. 
This is described by the non-linear
version of the MH surface diffusion equation with
$$
     A_{MBE}= -K \nabla^{4} h + \lambda_{1} \nabla^{2} (\nabla h)^{2}.
$$
The scaling exponents are known from a one-loop DRG calculation which
gives
$\alpha = (5-d)/3$, $\beta=(5-d)/(7+d)$ and $z=(7+d)/3$. 
There are two discrete models
to simulate the MBE universality class.

Lai-Das Sarma (LD) Model - This model is similar to the DT model 
described above 
with the difference that
if an atom falls in a kink site, it is allowed to break its bonds and jump
either down or up to the nearest kink site with the smallest 
step height \cite{laid}.

Kim-Das Sarma (KD) Model - This is the generalization of 
the KD model \cite{kim} 
described above. The only 
difference is that the linear curvature is replaced by a non-linear
curvature
=[h(x-1) + h(x+1) -2h(x)] - $\frac{\lambda}{2}$ \Big( $[h(x-1) - h(x)]^{2}$ +
$[h(x+1) - h(x)]^{2}$ \Big).  

\section{COMPUTER SIMULATIONS OF CYCLICAL GROWTH}

\subsection{Introduction} 
 
\subsubsection{Simulation methods} 

To test our scaling hypothesis for cyclical processes
we performed numerical simulations in d=1+1 using specific
discrete growth models described above.   
The system size in the simulations was changed between 128 to 4096 lattice
spacings. Periodic boundary condition is employed so that columns i
and i+L (L=system size) are equivalent. A typical
cycle consisted of a deposition of 5-20 layers (average number of particle
deposited per site) and desorption of
between 10$\%$ to 100$\%$ of the deposited amount. The maximum number of
cycles $n$
varied between $500 - 10000$ to reach the saturation. To obtain good
statistics we 
took average over 50-5000 independent runs,
depending on the pairs of primary processes
and the system size. The smaller the system the larger was the number of
runs to obtain good result.

\subsubsection{Time dependence of surface roughness} 

In a single process of
growth or desorption, the roughness scales with time (Eq. (2)). For cycling,
we have theoretically shown that time is replaced by the new scaling
variable: the number of cycles $n$. One might ask:  how does the cyclical
roughness change with actual time leading to a scaling in terms of $n$? 
Below we discuss how the time ($t$) dependence of the surface roughness of
cyclical growths can be studied to understand the emergent scaling
behavior in terms of the number of cycles $n$.

Consider a linear growth (or erosion) process. The structure factor for
this can be given by Eq. (11). From that expression of $S(q,t)$, the surface
width $W^{2} = 1/L^{\tilde d} \sum_{\vec{q}} S(\vec{q},t)$ can be expressed as \cite{krug}
\begin{equation}
W^{2}(t) = W_{o}^{2}(t) + W_{flat}^{2}(t),
\end{equation}
$W_{flat}(t)$ would be the roughness for growth induced on a 
flat initial substrate, and $W_{o}$ is the 
contribution due to the width of the rough substrate surface.
Since the total width is the sum of a decreasing ($W_{o}$) and
an increasing ($W_{flat}$) part,    
Competition between the two terms 
can make it increase or decrease from the initial roughness
$W_{o}(0)$ for sometime (eventually the roughness will exceed
$W_{o}(0)$). 

In cycling two primary processes act alternatingly. The roughness
generated by one process is taken as the initial roughness $W_{o}(0)$ for the 
second process. In Fig. 3 the height
profile is shown for the cyclic growth with Random deposition and EW
desorption. EW dissolution smoothes the very rough surface produced by
random growth in one cycle. Also note that the roughness increases 
with the number of cycles $n$.
In Fig. 4 we show the actual time dependence of the simulated 
cyclical growth composed of two linear primary processes belonging to
the EW and MH universality 
classes.
In one cycle, MH growth increases the
surface width and then EW surface relaxation smoothes out the surface
to lessen the roughness. The average behavior of surface roughness in
terms
of cycles increases and gives rise to scaling. The scaling exponents
are determined by the relevant (in the RG sense) terms of the 
primary processes. In our example (Fig. 4),
MH/EW cycles are dominated by the surface relaxation of EW model and
gives rise to EW exponents. The change in roughness
within a cycle can be thought of fluctuations to the average behavior, and
becomes less important as the number of cycles is increased (Fig. 4).
Similar scaling (in terms of $n$) continues to hold for cyclic 
growths when one or both the primary processes are non-linear.  

\subsubsection{Extraction of scaling exponents}

The growth exponent $\beta$ is extracted for different system sizes L. 
The roughness $W$ ${\it vs}$ $n$ is plotted on a log-scale and
the slope of the best fitted straight line yields the exponent $\beta$.
The value quoted is from the largest L (once the effective $\beta$ became 
close to the asymptotic value).
From $W_{s}(L) = W(L,\infty)$, the saturation width
dependence on $L$, the roughness exponent
$\alpha$ can be calculated. Simulation results for different system
sizes are used to plot $\ln W_{s} \sim ln L$, which is fitted to
a straight line whose slope measures the exponent $\alpha$. 
In some cases we checked independently
the value of $\alpha$ from the scale dependence of the height-difference
correlation function (from a log-log plot of $\Delta (r,t)$
${\it vs}$ $r$ and  fit to Eq. (4)).

\subsection{Simulation Results}

\subsubsection{Linear primary processes}

For linear primary processes,
we looked at the  possible pairwise combinations of RD, EW and MH 
universality classes using the 
absorption/desorption algorithms  
described earlier in this paper. When EW and MH is combined with RD, 
we obtain EW and MH exponents respectively, because those are the processes
which
generate correlations on top of the random growth.  Fig. 5 shows, for the
RD/EW cyclical process, that the 
value of $\beta$ (asymptotically) is
independent of the relative duration of the two primary processes. 
MH/EW cycles produce an asymptotic cyclical
scaling with EW exponents $\beta=0.258(5)$ (Fig. 6) and $\alpha=0.52(3)$
(Fig. 7). We have also performed data collapse (Fig. 8) to establish the
validity of our scaling hypothesis. Asymptotic EW exponents in MH/EW
cycles
confirms that the surface relaxation of EW is the dominating effect when 
paired with MH surface diffusion (or growth on kink sites). Our
theoretical
analysis also predicts that the EW scaling exponents will be imposed in
the
MH/EW cycles because EW has a smaller dynamic exponent ($z=2$) compared
to that ($z=4$) of the MH universality class.

\subsubsection{One non-linear process with a linear One} 

 To simulate non-linear processes we used two lattice realizations BD and
RSOS (with equivalent results) 
for the KPZ universality class and KD algorithm for the MBE universality. 
First we combine the KPZ universality with the EW or the MH 
universality in a cycle
to see the effect of the non-linear KPZ term to the 
scaling exponents. 
 The exponents obtained are $\beta=0.311(5)$,
 $\alpha=0.51(1)$ for EW/BD. The exponent $\beta$ increased slowly
with the system size and the effective $\beta$ reached a value close to
the asymptotic one only for the largest system size($L=4096$).
For the reverse process BD/EW we obtained and $\beta=0.322(5)$,
 $\alpha=0.50(1)$ (see Fig. 9). 
These asymptotic exponents are
 consistent with the KPZ $\beta=1/3$ and, of course, with
 $\alpha=1/2$,
 which is the common value of EW and KPZ.
To look at primary  processes with different values of $\alpha$, a
DT $(\alpha_{1}=1.5)$ deposition with a ballistic
desorption $(\alpha_{2}=0.5)$ were performed.
We obtain the asymptotic values of the
exponents for MH/KPZ:
$\beta=0.311(15)$ and $\alpha=0.48(2)$, both consistent with the KPZ
values. In all these cyclic processes the KPZ non-linearity 
$(\nabla h)^{2}$ remained relevant with respect to the linear terms 
($\nabla^{2} h$ or $\nabla^{4} h$)
and retain its scaling exponents. To show the opposite behavior where the 
non-linear process  is not the dominant one we simulated MBE/EW cycles.
From our approximate RG procedure for the cyclic process, it is clear that
the fourth-order non-linear perturbation $\nabla^{2}(\nabla h)^{2}$ 
is irrelevant with respect to the linear EW term $\nabla^{2} h$.
In our simulation,
when we allow surface relaxation of the EW model only to the nearest
neighbor
we get an effective exponent ($\beta \approx 0.31$) different than EW
$\beta$
even for the largest system size ($L=4096$) we used. The reason may be
that the next nearest neighbors of a chosen site  
affect the curvature dependent growth process in the simulation of MBE.
 When we consider next nearest neighbors for surface relaxation in EW
process
in MBE/EW cycles we get $\alpha=0.50(2)$ and $\beta=0.251(3)$ consistent
with EW values.

\subsubsection{Two non-linear processes}

Finally, we tried cycles consisting of two non-linear primary processes 
belonging to the KPZ and MBE universality classes. 
Simulations of RSOS/RSOS (note that
they are not time-reversed images of each other because of the
non-linearity)
gave surfaces with KPZ scaling for $T_{1} \neq T_{2}$. For $T_{1} =
T_{2}$,
however, EW behavior was found. This follows from the non-linear KPZ terms 
in the primary processes having the same magnitude but opposite signs.
Hence,
they exactly cancel each other in the coarse grained growth equations
yielding 
an EW one. When we combine KPZ (using BD and KK) with MBE (KD) we expect 
to get KPZ scaling for
cycles due to the dominant KPZ non-linearity (in d=1+1, where perturbative 
RG is applicable). Though both the processes have the same $\beta$ (in
1+1), 
in our 
simulation  of cycles we observe slow increase of beta with the system
size
when the BD model was used to simulate KPZ growth. 
The value of effective beta for the largest system size ($L=8192$) did not 
reach the
asymptotic value ($1/3$). 
Also, some sort of crossover behavior is observed
which varies depending on the ratio of $T_{1}$ and $T_{2}$.  The BD  model
itself 
is not very efficient to reach the asymptotic regime of KPZ growth and 
probably that is
reflected in the simulations of cyclical growth. When We used the RSOS
model for 
KPZ part of the cycle, the asymptotic scaling exponents ($\alpha=0.50(1)$
and
$\beta=0.335(5)$ ) were obtained for
relatively small system sizes. 

We also run the respective absorption/absorption (or growth/growth) cycles
with identical roughness behavior. Fig. 10 shows the results of DT/BD
growth/growth cycling simulation. The scaling exponents ($\beta =
0.31(2)$ and $\alpha = 0.48(2)$) again acquire the corresponding KPZ values
as expected.  

\section{EXPERIMENTS ON CYCLIC GROWTH}

Experiments of cyclical growth were performed by metal
electrodeposition/dissolution of silver. The scaling behavior of surface
roughness has been studied for electrochemical deposition 
\cite{iwamoto1} \cite{kaha} \cite{tong} \cite{schmidt} 
and dissolution \cite{iwamoto2} processes separately. In cycling, however,
metal electrodeposition, followed by a partial dissolution occurs. The
substrate is plated for a specific period of time and then the current is
reversed and the metal is dissolved from the substrate. This phenomenon is
inherent in rechargeable batteries.

Multiple cycles were carried out on vapor-deposited silver substrates,
ranging from $1$ to $20$ cycles. The plating solution contained $0.092M$
AgBr (silver bromide), $0.23M$ $(NH_{4})_{2}S_{2}O_{3}$ (ammonium
thiosulfate), and $0.17M$ $(NH_{4})_{2}SO_{3}$ (ammonium sulfite). Each
cycle consisted of plating for $5$ min followed by $2.5$ min of
electrodissolution with a current density of $0.8 mA/cm^{2}$. Image and
scaling analysis were done using an atomic force microscope (AFM).
Roughness was measured after $n$ full cycles and after the deposition part
of the cycles ( {\it i.e.} after $n+p$ cycles with $p=2/3$). 
A logarithmic plot of saturation rms height versus number of cycles, shown
in Fig. 11, resulted in a straight line over two decades, indicating that
the roughness scales with the number of cycles $n$, with $\beta=0.52$,
where $\beta$ is the growth exponent. The roughness for cycling and
deposition alone is compared. The value of $\beta$ is $0.52$ for the
cycling processes and $0.62$ for deposition alone \cite{dave}, 
suggesting that
processes such as the erosion of rough areas and the filling in of surface
recesses are occurring more during cycling than in straight deposition.
However, the saturation rms height was larger in magnitude for cycling
than for straight deposition.

The results suggest that cycling causes smoothing of the surface recesses
and valleys by the dissolution of large peaks during the reverse plating
process.

\section{CONCLUSIONS}

In this paper we have discussed $\it cyclical$ growth processes 
using dynamic scaling in terms of the number of cycles $n$ (replacing 
the time variable $t$ of a simple growth). Given two primary 
processes we have described how to derive the scaling exponents of the
combined cyclical process. MC simulations of different cyclic 
processes were carried out using various simple pairs of 
primary growth processes. Results of numerical simulations are entirely
consistent with our theoretical understanding. For linear processes,
the primary process with smaller dynamic 
exponent $z$ always imposes its own scaling exponents on the 
combined cyclical process. The application of the RG method to cyclical
processes was presented to determine cyclical scaling exponents
in the case of non-linear primary processes.
In particular, cyclical scaling exponents are independent
of the relative duration ($p$ or $1-p$) of the primary processes and 
also of the duration $T$ of the cycle. However, both $p$ and $T$ do
affect the amplitude of the roughness and also the crossover length
beyond which the asymptotic exponents show up.  
First experimental results on electrodeposition-
dissolution cycles are in accordance with our scaling hypothesis.

Cyclic growth phenomena are widespread in Nature: The sea shores
are shaped by cyclical high and low tides associated with the lunar 
motion around the Earth, all floral growth is subjected to daily
changes in light and seasonal variations (which also affect many
geological processes), etc. Thus the cyclical scaling theory 
introduced here might be found
helpful in future investigations of many natural growth phenomena.

Cyclical processes are also ubiquitous in technological applications.
For example corrosion processes are strongly affected by seasonal
weather conditions. The most likely application of our theory
is for rechargeable batteries. Indeed, one of the breakdown mechanisms
is due to the metal grown on one electrode reaching the other one
thereby causing a shortcircuit. To test batteries today one has to
run them their full lifetime. Using cyclical scaling, however,
accelerated testing will become feasible. Results from measurements
on fewer cycles for a short amount of time could be extrapolated
to predict the battery lifetime under its realistic working 
conditions.  

Finally application to medicine may also be envisioned. The first
one which comes to mind is the treatment of malignant tumors. They are
known to have rough surfaces which follow the kinetic growth scaling
laws \cite{tumor}. Under a chemotherapeutic (or radiation) treatment 
they shrink and,
hopefully, disappear. These treatments are always cyclical, and thus
the cyclical scaling approach should faithfully describe the tumor's
recession. We hope that using this approach will help plan the schedule
and the dosage of the treatments such as to make them more efficient 
while minimizing the side effects.

\section{ACKNOWLEDGMENTS}
One of us (S. R.) is grateful to S. Raghavan for his help in using the
software Xfig. This work was supported by the ONR grant 
N00014-00-1-0057.

\begin{figure}
\centerline{\epsfig{file=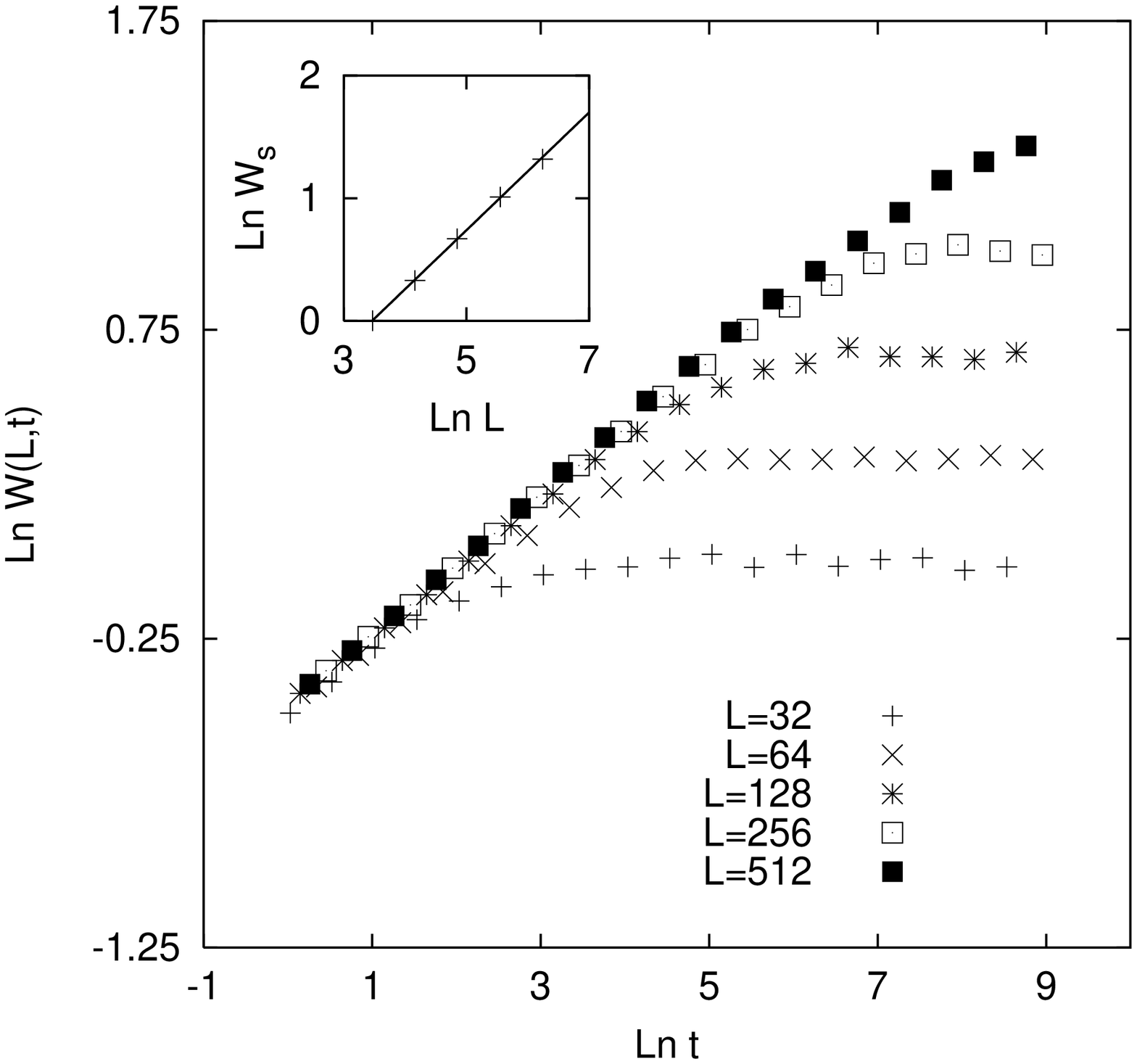,height=3.5in,width=3.5in,angle=-90}}
\vspace{0.5cm}
\caption{The roughness $W$ {\it vs}  $t$ in the EW desorption process 
({\it log-log} plot). (Inset: saturation roughness $\ln W_{s}$ 
{\it vs} $\ln L$).} 
\label{Fig. 1}
\end{figure}

\vspace{0.8in}

\begin{figure}
\centerline{\epsfig{file=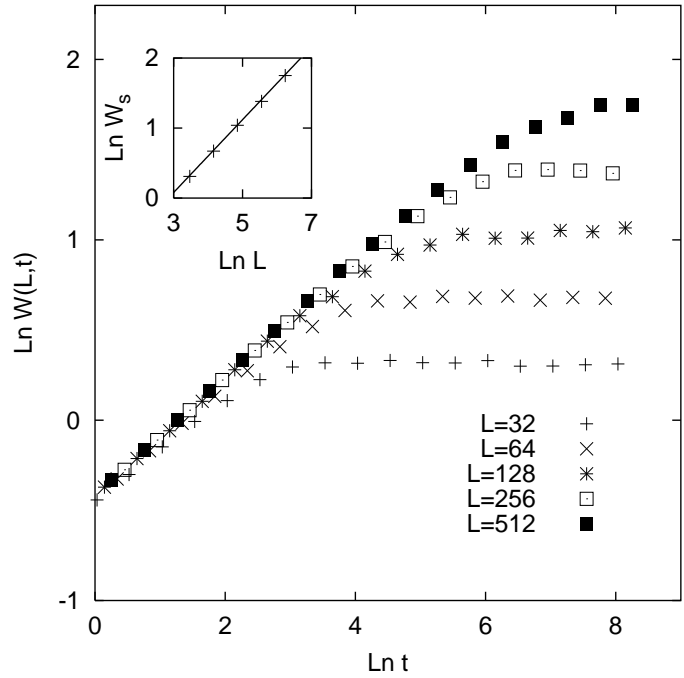,height=3.5in,width=3.5in,angle=-90}}
\vspace{0.5cm}
\caption{The roughness $W$ {\it vs}  $t$ in the KPZ (RSOS)  desorption
process ({\it log-log} plot). (Inset: maximal roughness $\ln W_{s}$ 
{\it vs} $\ln L$).}
\label{Fig. 2}
\end{figure}

\begin{figure}
\centerline{\epsfig{file=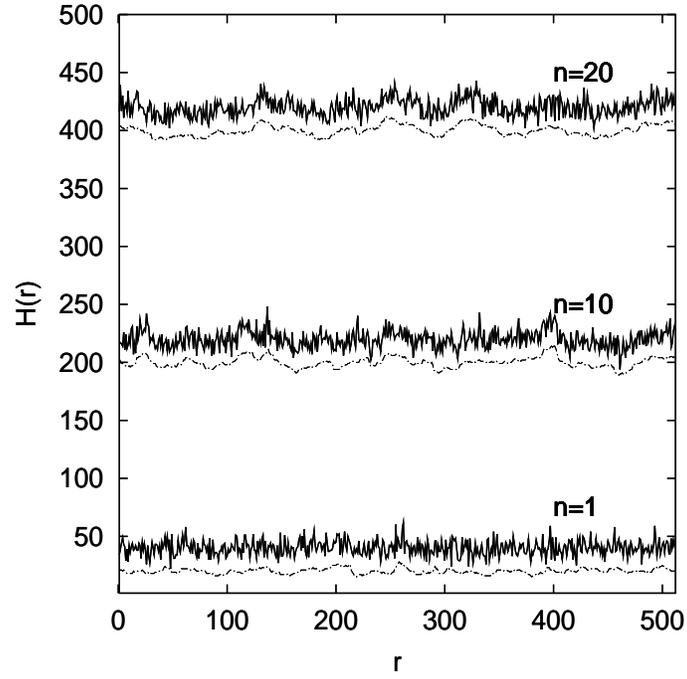,height=3.5in,width=3.5in,angle=-90}}
\vspace{0.5cm}
\caption{Height profile for RD/EW cyclical growth. L=512 and 
40 particles/site of Random deposition (solid line) and 
20 particles/site of EW dissolution (broken line) are used in one cycle.} 
\label{Fig. 3}
\end{figure}

\vspace{0.8in}

\begin{figure}
\centerline{\epsfig{file=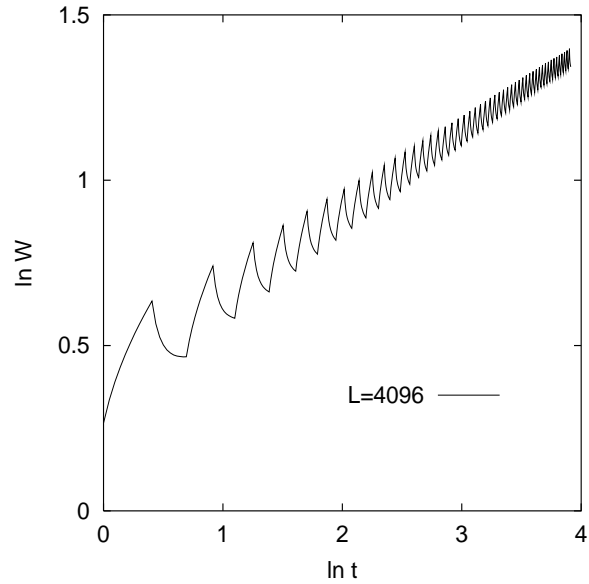,height=3.in,width=3.in,angle=-90}}
\vspace{0.5cm}
\caption{$\ln W$ (roughness) $\sim$ $\ln t$ of the MH/EW cyclic process. DT process increases the roughness
and EW smoothes the surface in a single cycle, but the average roughness 
increases and gives rise to scaling (in terms of cycles).}
\label{Fig. 4}
\end{figure}  
 
\begin{figure}
\centerline{\epsfig{file=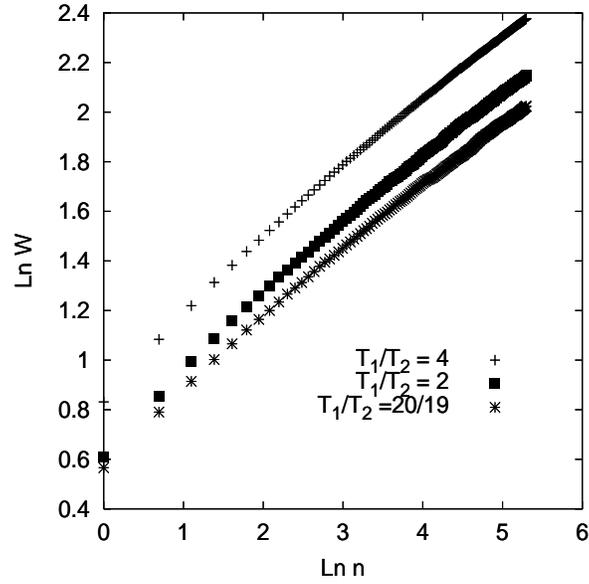,height=3.in,width=3.in,angle=-90}}
\vspace{0.5cm}
\caption{The roughness $W$ {\it vs} number of cycles $n$ in the
RD/EW cyclical growth with different ratios of deposition $T_{1}$
and dissolution $T_{2}$ ({\it log-log} plot). System size of L=1024
and deposition of $20000$ particles (fixed) are used.}
\label{Fig. 5}
\end{figure}

\vspace{0.8in}

\begin{figure}
\centerline{\epsfig{file=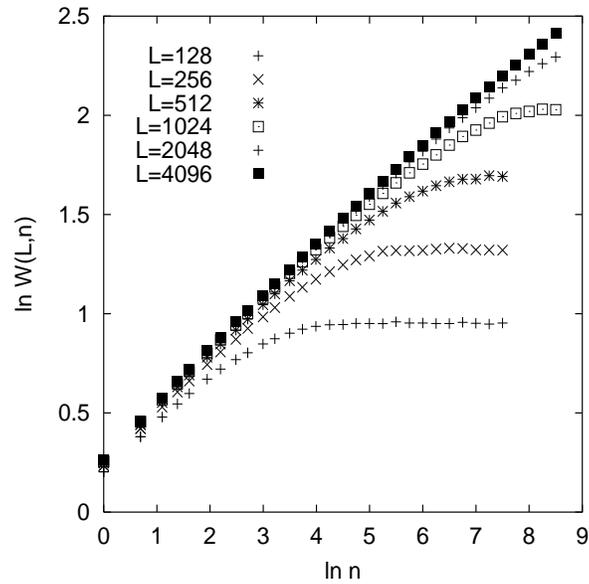,height=3.in,width=3.in,angle=-90}}
\vspace{0.5cm}
\caption{$\ln W$ (roughness) {\it vs}
$\ln n$ (number of cycles) of the MH/EW cyclical process 
for different system sizes $L$.} 
\label{Fig. 6}
\end{figure}

\begin{figure}
\centerline{\epsfig{file=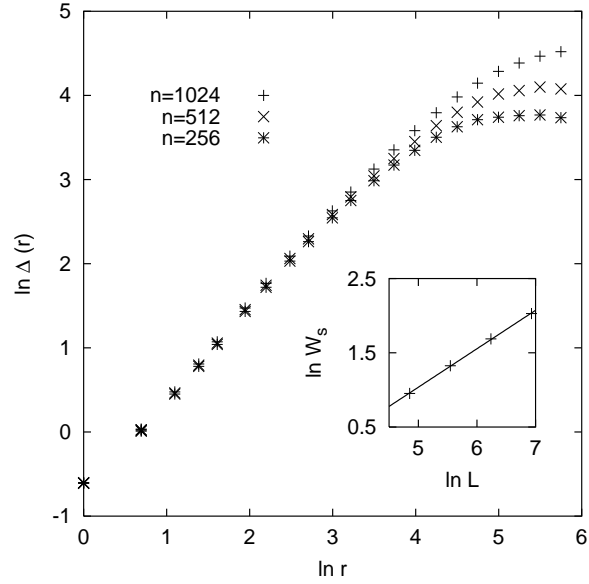,height=3.in,width=3.in,angle=-90}}
\vspace{0.5cm}
\caption{Height-difference correlation function $\Delta(r)$ {\it vs} distance
$r$
for the MH/EW cyclic growth 
is plotted after different number of cycles ({\it log-log} plot). 
Inset: $\ln{W_{s}}$ (maximal roughness calculated from Fig. 6) {\it vs} 
$\ln{L}$.} 
\label{Fig. 7}
\end{figure}

\vspace{0.8in}

\begin{figure}
\centerline{\epsfig{file=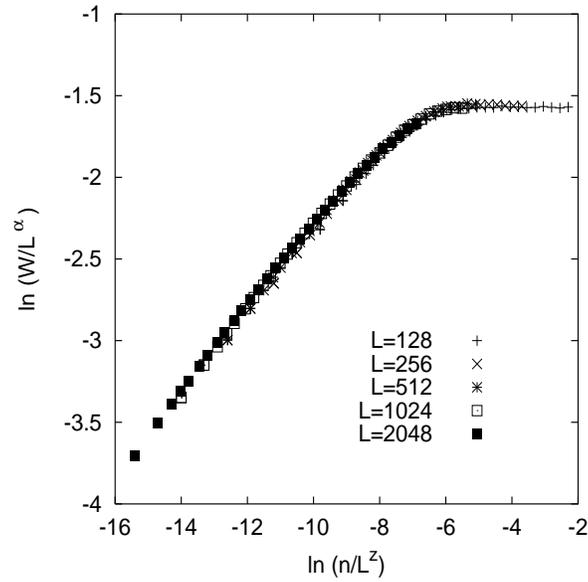,height=3.in,width=3.in,angle=-90}}
\vspace{0.5cm}
\caption{Data collapse for MH/EW cyclical growth (using the 
exponents found from the graphs in
Fig.4 and Fig.5.) clearly shows scaling in terms of cycles $n$.}
\label{Fig. 8}
\end{figure}

\begin{figure}
\centerline{\epsfig{file=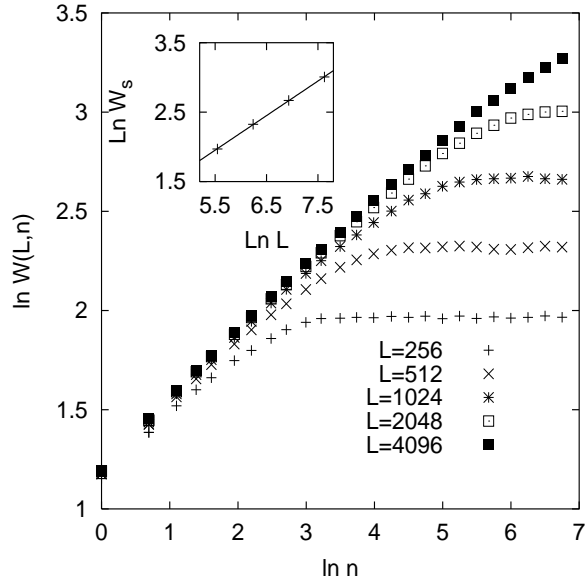,height=3.in,width=3.in,angle=-90}}
\vspace{0.5cm}
\caption{$\ln W$ (roughness) {\it vs}
$\ln n$ (number of cycles) of the simulated KPZ/EW cyclic growth for 
different system sizes L. Inset:roughness exponent $\alpha$ is extracted 
from its maximal values $W_{s}$ for different $L$.}
\label{Fig. 9}
\end{figure}

\vspace{0.8in}

\begin{figure}
\centerline{\epsfig{file=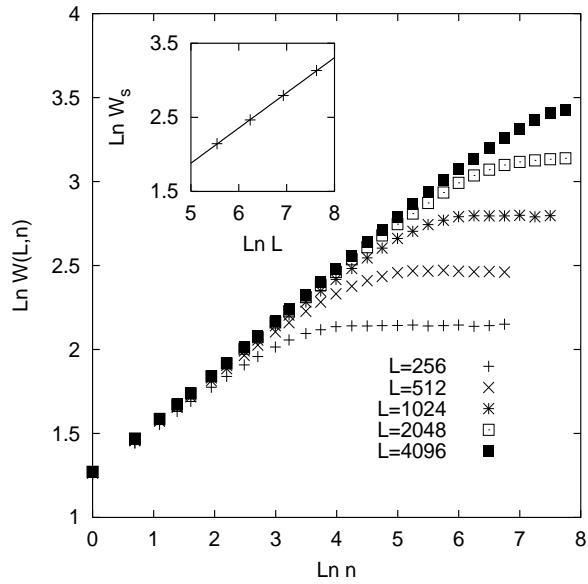,height=3.in,width=3.in,angle=-90}}
\vspace{0.5cm}
\caption{The roughness $W$ {\it vs} $n$  
of the DT/BD (growth/gowth) cyclical process ({\it log-log} plot).
Inset: saturation roughness $\ln{W_{s}}$ {\it vs} $\ln{L}$.} 
\label{Fig. 10}
\end{figure}

\begin{figure}
\centerline{\epsfig{file=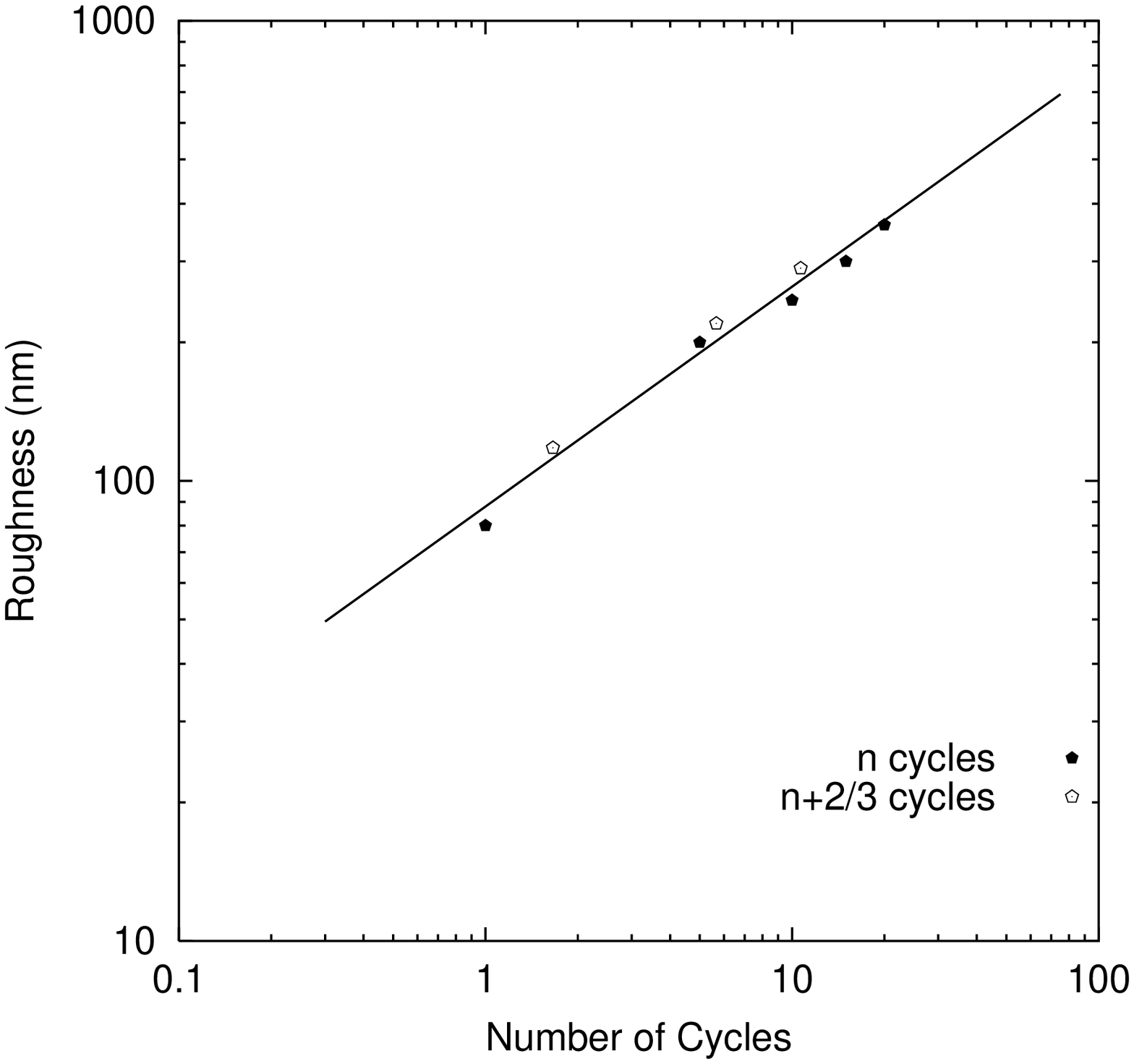,height=3.in,width=3.in,angle=-90}}
\vspace{0.5cm}
\caption{The roughness {\it vs} number of cycles $n$ in the
electrochemical cyclical growth of Silver ({\it log-log} plot).}
\label{Fig. 11}
\end{figure}


\begin{references}

\bibitem{family1}
{\it Dynamics of Fractal Surfaces}, edited by F. Family and T.
Vicsek (Cambridge University Press, Cambridge, England, 1990).

\bibitem{bara}
A.-L. Barabasi and H. E. Stanley, {\it Fractal Concepts in Surfaces
Growth}, (Cambridge University Press, Cambridge, England, 1995).

\bibitem{villain} 
J. Villain and A. Pimpinelli, {\it Physique de la Croissance
Crystalline}, (Editions Eyrolles, Paris, France, 1995).

\bibitem{zhang}
T. Halpin-Healy and Y.-C. Zhang, Phys. Rep. {\bf 254}, 215 (1995).

\bibitem{meakin1}
P. Meakin, {\it Fractals, scaling and growth far from equilibrium},
(Cambridge University Press, Cambridge, England, 1998)

\bibitem{pelletier}
J. D. Pelletier, ao-sci/9605001.

\bibitem{burgers}
J. M. Burgers, {\it The nonlinear diffusion equation: asymptotic 
solutions and statistical problems}, (D. Riedel Pub. Co., Boston, 
1974). 

\bibitem{zia}
G. Foltin, K. Oerding, Z. Racz, R. L. Workman, and R. K. P. Zia,
Phys. Rev. E {\bf 50}, R639, (1994).

\bibitem{plis}
Zoltan Racz and Michael Plischke, Phys. Rev. E {\bf 50}, 3530, (1994).

\bibitem{bray}
J. M. Kim, M. A. Moore, and A. J. Bray, Phys. Rev. A {\bf 44},
2345 (1991).

\bibitem{derrida}
B. Derrida and J. L. Lebowitz, Phys. Rev. Lett. {\bf 80}, 209, 1998;
B. Derrida and C. Apert, J. of Stat. Phys. {\bf 94}, 1, 1999.

\bibitem{ziadasarma}
 Z. Toroczkai, G. Korniss, S. Das Sarma, and R. K. P. Zia,
Phys. Rev. E {\bf 62}, 276, (2000).

\bibitem{PhysicsToday}
Z. Toroczkai and E. D. Williams, Physics Today, 24, December (1999);
J. Krug et al. Phys. Rev. E {\bf 56}, 2702, (1997); H. Kallabis and
J. Krug, cond-mat 9809241; S. N. Majumdar, Curr. Sci. {\bf77}, 370  
(1999).

\bibitem{subha}
S. Raychaudhuri, M. Cranston, C. Przybyla, and Y. Shapir, 
cond-mat/0105176.

\bibitem{iwamoto1}
A. Iwamoto, T. Yoshinobu, and H. Iwasaki, Phys. Rev. Lett. 
{\bf 72}, 4025 (1994).

\bibitem{tumor}
A. Bru {\it et al}, Phys. Rev. Lett. {\bf 81}, 4008 (1998).

\bibitem{iwamoto2}
A. Iwamoto, T. Yoshinobu, and H. Iwasaki, Phys. Rev. E
{\bf59}, 5133 (1999).

\bibitem{family2}
 F. Family and T. Vicsek, J. Phys. A{\bf 18}, L75 (1985).

\bibitem{shapir}
Y. Shapir, S. Raychaudhuri, D. G. Foster, and J. Jorne, Phys. Rev.
Lett. {\bf 84}, 3029 (2000).

\bibitem{dkim}
M. Kardar (p. 30) and B. Kahng (p. 67) in 
{\it Dynamics of Fluctuating Interfaces and Related Phenomena},
edited by D. Kim, H. Park, and B. Kahng (World Scientific, Singapore,
1997).

\bibitem{ew}
S. F. Edwards and D. R. Wilkinson, Proc. Roy. Soc. London A {\bf 381},
17 (1982).

\bibitem{family3}
F. Family, J. Phys. A{\bf 19}, L441 (1986).

\bibitem{kardar}
M. Kardar, G. Parisi, and Y.-C. Zhang, Phys. Rev. Lett. {\bf 56},
889 (1986).

\bibitem{vold}
M. J. Vold, J. Coll. Sci. {\bf 14}, 168 (1959).

\bibitem{meakin2}
P. Meakin, P. Ramanlal, L. M. Sander, and R. C. Ball, Phys. Rev.
A{\bf 34}, 5091 (1986).

\bibitem{kost}
J. M. Kim and J. M. Kosterlitz, Phys. Rev. Lett. {\bf 62}, 2289 (1989).

\bibitem{MH}
W. W. Mullins, J. Appl. Phys. {\bf 28}, 333 (1957); C. Herring, J. Appl.
Phys. {\bf 21}, 301 (1950) 

\bibitem{dasarma1}
S. Das Sarma and P. Tamborenea, Phys. Rev. Lett. {\bf 66}, 325 (1990).

\bibitem{wolf}
D. E. Wolf and J. Villain, Europhy. Lett. {\bf 13}, 389 (1990).

\bibitem{dasarma2}
S. Das Sarma, cond-mat/9705118 and references therein.

\bibitem{kim}
J. M. Kim and S. Das Sarma, Phys. Rev. Let. {\bf 72}, 2903 (1994).

\bibitem{laid}
Z. W. Lai and S. Das Sarma, Phys. Rev. Lett. {\bf 66}, 2348 (1991)

\bibitem{vill}
J. Villain, J. Phys. (france) {\bf 1}, 19 (1991)

\bibitem{krug}
S. Majaniemi, T. Ala-Nissila, and J. Krug, Phys. Rev. B. {\bf 53},
8071 (1996); J. Krug, Adv. in Phys. {\bf 46}, 139 (1997).

\bibitem{kaha}
G. L. M. K. S. Kahanda, X. Zou, R. Farrel, and P.Wong, Phys. Rev. Lett.
{\bf 68}, 3741 (1992).

\bibitem{tong}
W .M. Tong and R. S. Williams, Annu. Rev. Phys. Chem.
{\bf 45}, 401 (1994).

\bibitem{schmidt}
W. U. Schmidt, R. C. Alkire, and A. A. Gewirth, J. Electrochem. Soc.,
{\bf 143}, 3122 (1996).

\bibitem{dave}
D. G. Foster, Ph.D. Thesis, University of Rochester (1999);
D. G. Foster, J. Jorne, Y. Shapir, and S. Raychaudhuri,
Abs. {\bf 844}, Proceedings of ``The 196th Meeting of The
Electrochemical Society, Inc.'', Hawaii (1999).

\end{references}
\end{document}